\documentclass[prb,twocolumn,showpacs]{revtex4}
\usepackage{graphicx}
\usepackage{dcolumn}

\begin{document}

\title{Subgap states in dirty superconductors and their effect on dephasing
in Josephson qubits }
\author{Alessandro Silva and Lev B. Ioffe}
\affiliation{Department of Physics and Astronomy, Rutgers University, 136 Frelinghuysen
Road, Piscataway 08854, New Jersey, USA.}
\date{\today }

\begin{abstract}
We present a theory of the subgap tails of the density of states in a 
diffusive superconductor containing magnetic impurities. We show that
the subgap tails have two contributions: one
arising from mesoscopic gap fluctuations, previously discussed by
Lamacraft and Simons, and the other associated to the 
long-wave fluctuations of the concentration of magnetic impurities. 
We study the latter both in small superconducting grains and in bulk systems 
[$d=1,2,3$], and establish the dimensionless parameter that controls 
which of the two contributions dominates the subgap tails.  
We observe that these contributions are related to each other by 
dimensional reduction. 
We apply the theory to estimate the effects of a weak concentration
of magnetic impurities [$\approx 1 {\rm p.p.m}$] on the phase coherence
of Josephson qubits. We find that at these typical concentrations, 
magnetic impurities are relevant for the dephasing in 
large qubits, designed around a $10\;{\rm \mu m}$ scale, 
where they limit the quality factor to be $Q<10^4-10^5$.    
\end{abstract}


\maketitle



\section{Introduction.}

\label{sec0}

Recent experimental studies of disordered conductors demonstrated that
even a weak concentration of magnetic impurities may lead to important
effects, especially in the context of dephasing and energy relaxation. In
particular, the experiments in mesoscopic Au and Cu wires have shown that
the frequently observed saturation of the dephasing time $\tau _{\varphi }$
at low temperatures~\cite{Mohanty, Pierre, Pierre2} is likely due to a
small unavoidable presence of magnetic impurities with the concentration as
low as $0.1-1\;{ppm}$. On a theoretical side it was shown~\cite{Glazman}
that magnetic impurities at these concentrations can explain the puzzling
anomalous energy dependence of the relaxation rate observed in diffusive Au
and Cu wires~\cite{Pothier,Pierre3}.

It is well known that a significant concentration of magnetic impurities
strongly affects conventional superconductors~\cite{Abrikosov}. Further, it
was shown recently that even a small concentration of magnetic impurities
leads to exponentially small tails in the density of states within the
gap ~\cite{Balatsky, Aleiner,Shytov, Brouwer, Lamacraft}. Because such
subgap states can trap and release the quasiparticles, they can become an
important for a number of physical applications, e.g. for Josephson-junction
qubits. For example, since the resulting subgap states are localized, the
quasiparticles that are generated during qubit manipulations or readout ~%
\cite{Martinis1}, can get trapped in these localized subgap states. These
trapped quasiparticles provide extra degrees of freedom that contribute to
the dephasing and dissipation in subsequent qubit operations. Although the
typical materials used for the fabrication of Josephson-junction qubits [
e.g., ${Al-Al_{2}O_{3}-Al}$], are not intentionally doped with magnetic
impurities, some concentration of them is unavoidable, especially in small
devices where they are due to the surface effects. Therefore, it is
important to estimate the effects of diluted magnetic impurities
on the density of states in these systems and the resulting dephasing.

With this goal in mind we present here a detailed theory of the subgap
states in a diffusive superconductors containing low concentrations of
magnetic impurities. In particular, we study the contribution to the subgap
density of states due to spatial fluctuations of the concentration of
magnetic impurities. Combining the results of our analysis
with those recently obtained by
Lamacraft and Simons~\cite{Lamacraft}, who considered the subgap tails
resulting from universal mesoscopic gap fluctuations, we provide a full
picture of the subgap states, and establish in which regime one or the other
type of the fluctuation dominates the physics. Finally, the results of our
theoretical analysis are applied to estimate the subgap density of states
generated by a concentration of $\approx 1$ ppm of magnetic impurities in
aluminum-based qubits. The result of our analysis is that the smallest
superconducting islands used in modern experiments might contain a few
subgap states while larger qubits can contain about $10^{3}$ of them.
Finally, we give the estimates the qubit decoherence resulting from trapped
quasiparticles and conclude that they provide an important source of the
decoherence in larger devices but negligible in the smallest ones.

The rest of paper is organized as follows. In Sec.~\ref{sec00} we present
the qualitative arguments and the results of our analysis. The technical
details of the density of states computation are presented in Sec.~\ref%
{sec01}: there we first discuss the effect of fluctuations on the disorder
averaged density of states in a large collection of superconducting grains
[~Sec.~\ref{sec1}], and then extend the analysis to the bulk systems ($d=1,2,3$%
) in ~Sec.~\ref{sec2}. In Sec.~\ref{sec2b} we present the details of our
estimates of the subgap density of states in superconducting qubits and its
effect on the decoherence. Finally, Sec.~\ref{sec3} gives the conclusions.

\section{Overview and Main Results.}

~\label{sec00}

The effects of impurities on conventional superconductivity were
extensively studied in the last four decades. It is well known that
dilute non-magnetic impurities preserving the time reversal symmetry do not
affect the superconducting gap \cite{Anderson} while a finite concentration
of magnetic impurities suppresses both the critical temperature $T_{c}$ and
the gap in the local density of states, $E_{g}$ \cite{Abrikosov}. The effect
of weak magnetic impurities is controlled by the dimensionless parameter 
\begin{equation}
\zeta =\frac{1}{\tau _{s}\Delta },
\end{equation}%
where $\tau _{s}$ is the spin-scattering time and $\Delta $ the bare
superconducting gap. At the critical concentration of magnetic impurities
corresponding to $\zeta =1$ the gap in the density of states closes but $%
T_{c}$ remains finite; at larger concentrations one the superconductor
becomes gapless~\cite{Abrikosov}. Subsequent extensions of this theory have
included the effect of strong magnetic impurities. While a single strong
magnetic impurity generates a bound intra-gap quasi-particle state and a
local suppression of the BCS gap~\cite{Liu,Rusinov,Balatsky3}, a finite
concentration of them leads to the formation of an intra-gap impurity band~%
\cite{Liu,Shiba,Rusinov2}. As the concentration is further increased, the
impurity band broadens, merging with the BCS continuum, and eventually
resulting in a complete gap destruction.

For small impurity concentration corresponding to $\zeta <1$, the
Abrikosov-Gorkov theory \cite{Abrikosov} predicts a vanishing density of
states for $\mid E\mid <E_{g}=\Delta (1-\zeta ^{2/3})^{3/2}$. This
conclusion, however, neglects the effects of the impurity distribution
fluctuations \cite{Balatsky}. One expects that such fluctuations smear the
gap edge and lead to the exponential tails in the density of states
extending down into the superconducting gap, similarly to the Lifshits tails~%
\cite{Lifshits} in the density of states below the band edge of a disordered
conductor~\cite{Halperin,Lax,Zittarz}.

As in the case of Lifshits tails one can visualize the the subgap states in
a bulk sample as appearing locally in places where the impurity potential
decreases the energy of the quasiparticle. Further, the main contribution to
the states deep in the tail is dominated by the specific impurity potential
that is called the optimal fluctuation. Historically, the first example of
such optimal fluctuations in the context of superconductivity was found by
Larkin and Ovchnikov~\cite{Larkin} who considered the effects of the 
interaction constant inhomogeneity on the density of the subgap states.  
Recently, various mechanisms leading to
such sub-gap states in superconductors containing magnetic impurities have
been considered, both in lightly magnetically doped superconductors~\cite%
{Shytov}, in small superconducting grains~\cite{Aleiner, Brouwer,Lamacraft},
and in diffusive superconductors ~\cite{Balatsky, Lamacraft}. We will
concentrate on the latter case: dirty superconductors containing both
magnetic and non-magnetic impurities, under the conditions $\zeta \ll 1$ and 
$l\ll \xi $, where $l$ is the mean free path relative to momentum scattering
and $\xi $ is the coherence length.

The subgap tails in this regime were recently considered by Lamacraft and
Simons~\cite{Lamacraft}. Slightly above the gap the density of states
predicted by Abrikosov and Gorkov is 
\begin{eqnarray}
&&  \nonumber \\
\nu &=&\frac{1}{\pi L^{d}}\sqrt{\frac{E-E_{g}(\zeta )}{\Delta _{g}^{3}}}
\label{AG_DOS} \\
\Delta _{g}^{-3/2} &\simeq & \pi \nu _{0}L^{d}\sqrt{\frac{2}{3\Delta }}\frac{%
1}{\zeta ^{2/3}}
\end{eqnarray}%
where $\nu _{0}$ is the density of states per unit volume in the normal
state, $L$ is the sample size, and $d$ the dimensionality of the system.
Formulating the problem in terms of a supersymmetric Sigma model, Lamacraft
and Simons obtained a density of states close but below the gap 
\begin{equation}
\frac{\langle \nu \rangle }{\nu _{0}}\propto \exp \left[ -\tilde{a}%
_{d}\left( \frac{\lambda _{0}}{L}\right) ^{d}\left( \frac{E_{g}-E}{\Delta
_{g}}\right) ^{3/2}\right]  \label{Lama}
\end{equation}%
where $\tilde{a}_{d}\sim 1$ and 
\begin{equation}
\lambda _{0}=\xi \left( \frac{\Delta }{E_{g}-E}\right) ^{1/4}  \label{size}
\end{equation}%
is the linear size of the optimal fluctuation.

This result has a clear physical interpretation. In zero dimensions, the
action is a universal function of the rescaled energy $(E_{g}-E)/\Delta_g$,
where $\Delta_g$ can be interpreted as the effective level spacing in the
right above the AG gap. This indicates that the gap fluctuations can be seen
as Random Matrix-like fluctuations of the edge of a Wigner semicircle~\cite%
{Tracy}. Indeed, the zero dimensional result was previously conjectured~\cite%
{Brouwer} on the basis of the universality hypothesis of Random Matrix
Theory.

In a bulk system [$d\neq 0$], the subgap localized states giving the density
of states Eq.(\ref{Lama}) originate from particular configurations of normal
and magnetic impurities favoring mesoscopic fluctuations of the gap edge. On
the other hand, it is intuitively clear that another type of inhomogeneous
fluctuations should also contribute to the sub-gap tails of the DOS: local
fluctuations of the concentration $n_{imp}$ of magnetic impurities dictated
by their Poissonian statistics . A local increase of $n_{imp}$, and
therefore of the spin-scattering rate $\zeta $, above its average value
implies a local suppression of the AG gap, and as a result the formation of
sub-gap localized quasi-particle states~\cite{Balatsky1}.

Below, we will show that the mechanism generating subgap states described
above is complementary to universal mesoscopic gap fluctuations. This
complementarity is particularly transparent in the peculiar zero dimensional
limit, where the explicit form of the subgap tails depends on the way the
statistics over disorder realizations is acquired . Indeed, if gap
fluctuations are studied in a single sample, e.g varying the boundary
conditions through gate voltages, the rate $\zeta $ is constant, and the
universal result of Lamacraft and Simons~\cite{Lamacraft} always applies. On
the other hand, if one considers a large collection of the superconducting
grains $\zeta $ is going to display sample to sample fluctuations. As shown
below, depending on the system parameters, either the former or the latter
effect is bigger.

While in the zero dimensional case one has the possibility to single out
mesoscopic gap fluctuations by considering a single mesoscopic grain, in
finite dimensional systems this is not possible. Since the analysis of Ref.[%
\onlinecite{Lamacraft}] neglects large scale fluctuations of $\zeta $,
limiting the scattering rate to be a constant independent on position, only
a direct comparison of the resulting subgap tails can establish which of the
two mechanisms gives the dominant contribution to the asymptotic subgap
density of states.

In the following Sections, we shall study in detail the physics of the
subgap states due to the fluctuations of the concentration of magnetic
impurities, or, equivalently, of $\zeta $. We describe the dirty
superconductor by the quasiclassical Usadel equations~\cite{Usadel,Kopnin}
in which the spin-scattering rate $\zeta $ becomes a position dependent
statistical quantity. The variable $\zeta $ inherits its statistics from the
Poissonian distribution of magnetic impurities. In this framework, we find
the optimal fluctuation of $\zeta $ by solving the associated variational
problem and the resulting instanton equations, and discuss the deep
analogies between the problem at hand and the Lifshits tails in disordered
conductors. We then calculate the expression of the asymptotic subgap tails
of the DOS, including gaussian prefactors, as a function of the distance
from the gap edge 
\[
\delta \epsilon =\frac{(E_{g}-E)}{\Delta} , 
\]%
and of the average scattering rate $\zeta $, impurity concentration $n_{imp}$%
, and dimensionality [$d=0,1,2,3$]. For every dimensionality, we compare our
results with the subgap tails due to mesoscopic gap fluctuations and
establish the parameter that controls which of the two types of fluctuations
dominated the physics of subgap states.

The results of our analysis may be summarized as follows. While the typical
size of the optimal fluctuations associated to a local increase of $\zeta $
is given by Eq.(\ref{size}), the subgap density of states originating from
the $n_{imp}$ fluctuations is \ 
\begin{eqnarray}\label{DOS0}
\frac{\langle \nu \rangle }{\nu _{0}} &\propto &\exp \left[ -a_{d}\frac{%
n_{imp}\xi ^{d}}{\zeta ^{4/3}}(\delta \epsilon )^{2-d/4}\right] \nonumber 
\\
&=&\exp \left[ -a_{d}\left( \frac{\lambda _{0}}{L}\right) ^{d}\left( \frac{%
E_{g}-E}{\delta E_{g}}\right) ^{2}\right]
\end{eqnarray}%
where $a_{d}$ is a constant, calculated below, $\delta E_{g}=\Delta \zeta
^{2/3}/\sqrt{N}$, and $N=n_{imp}\;L^{d}$. A more detailed formula for the
DOS is given by Eq.(\ref{finalDOS}). The effective dimensionality of the system
is determined by comparing the linear size of the sample to the typical size
of an optimal fluctuation $\lambda $, Eq.(\ref{size}). In particular, for $d=0$
the parameter that determines whether the physics of tail states is
dominated by mesoscopic fluctuations [Eq.~(\ref{Lama})] or fluctuations of $%
\zeta $ [Eq.~(\ref{DOS0})] is the ratio 
\[
\beta =\frac{\Delta _{g}}{\delta E_{g}} 
\]%
For $\beta \gg 1$ the asymptotics is dictated by Eq.(\ref{Lama}), while for $%
\beta \ll 1$ the result of Eq.(\ref{DOS0})
applies. In higher dimensions, a detailed comparison
is more involved. However it is interesting to notice that, apart from a
numerical constant, the ratio of the actions relative to the two optimal
fluctuations [see Eq.~(\ref{Lama}), and Eq.~(\ref{DOS0})] is independent of
dimensionality.

\section{Subgap DOS associated to fluctuations of $\protect\zeta$.}

~\label{sec01}

In this Section, we present the theory of the subgap density of states
associated to fluctuations of the spin-scattering rate $\zeta $, starting
with the simple zero dimensional case, and subsequently extending it to the
case of finite dimensional systems.

\subsection{Fluctuations of $\protect\zeta$ in small grains.}

~\label{sec1}

We begin with the simpler zero dimensional case, the results in this
subsection apply to the samples which are smaller than the size of the
optimal fluctuation.

As explained in the previous section, the zero dimensional case is peculiar
because the form of the subgap DOS depends on the whether statistics is
acquired measuring the DOS in a single sample, e.g. varying the boundary
conditions, or in a large collection of small grains. In the first case, the
subgap tails are always dominated by mesoscopic Random Matrix-like gap
fluctuations. In the following, we will analyze the case of a large
collection of small grains, where both mesoscopic gap fluctuations and
fluctuations of $\zeta $ contribute to the subgap tails.

For every value of the dimensionless spin-scattering rate $\tilde{\zeta}$,
and slightly above the gap edge, the DOS in the Abrikosov-Gorkov
approximation is given by 
\begin{eqnarray}
\nu &=&\frac{1}{\pi L^{3}}\sqrt{\frac{E-E_{g}(\tilde{\zeta})}{\Delta _{g}^{3}%
}}, \\
\Delta _{g}^{-3/2} &\simeq & \pi \nu _{0}L^{3}\sqrt{\frac{2}{3\Delta }}\frac{%
1}{\tilde{\zeta}^{2/3}}.  \label{levspacing}
\end{eqnarray}%
where $L$ is the linear size of each grain. We can estimate the DOS tail resulting from
sample to sample fluctuations of the concentration of impurities by
promoting $\tilde{\zeta}$ to be a statistical variable, with average $\zeta $
and variance $\zeta ^{2}/N$, where $N$ is the average number of paramagnetic
impurities per sample. The average DOS is 
\[
\frac{\langle \nu \rangle }{\nu _{0}}=\frac{4}{\zeta ^{2/3}}\sqrt{\frac{2}{3}%
}\int \frac{d(\delta \zeta )}{\sqrt{2\pi \eta _{0}}}\sqrt{\frac{%
E-E_{g}(\zeta +\delta \zeta )}{\Delta }}e^{-\frac{(\delta \zeta )^{2}}{2\eta
_{0}}}, 
\]%
where $\eta _{0}=\zeta ^{2}/N$. Since 
\[
\frac{E_{g}(\zeta +\delta \zeta )}{\Delta }\simeq \frac{E_{g}(\zeta )}{%
\Delta }-\frac{\delta \zeta }{\zeta ^{1/3}}, 
\]%
introducing $\delta \epsilon =(E_{g}(\zeta )-E)/\Delta $, we obtain 
\begin{eqnarray}
\frac{\langle \nu \rangle }{\nu _{0}} &\propto &\int_{\zeta ^{1/3}\delta
\epsilon }^{+\infty }d(\delta \zeta )\sqrt{\delta \zeta -\zeta ^{1/3}\delta
\epsilon }\;e^{-\frac{(\delta \zeta )^{2}}{2\eta _{0}}}  \nonumber
\label{finalzeroDOS} \\
&\propto &\left[ \frac{1}{2S}\right] ^{3/4}e^{-S}.
\end{eqnarray}%
Here 
\begin{equation}
S=\frac{1}{2}\frac{N}{\zeta ^{4/3}}(\delta \epsilon )^{2}=\frac{1}{2}\left( 
\frac{E_{g}-E}{\delta E_{g}}\right) ^{2},  \label{actionme}
\end{equation}%
is the action, assumed to be $S>>1$, and 
\[
\delta E_{g}=\Delta \frac{\zeta ^{2/3}}{\sqrt{N}}, 
\]%
is the typical scale of gap fluctuations. The final result is 
\[
\frac{\langle \nu \rangle }{\nu _{0}}\propto \frac{1}{(\delta \epsilon
)^{3/2}}\exp \left[ -\frac{1}{2}\frac{N}{\zeta ^{4/3}}(\delta \epsilon )^{2}%
\right] . 
\]%

Let us compare now these results to the asymptotic DOS resulting from
mesoscopic fluctuations of the gap edge~\cite{Lamacraft}. These fluctuations
are characterized by the action 
\[
S_{SUSY}=\frac{4}{3}\left( \frac{E_{g}-E}{\Delta _{g}}\right) ^{3/2}. 
\]%
As consequence of the universality of this result, this action is of order $%
1 $ at energies below the gap corresponding to the effective level spacing
right above the AG gap edge, i.e. 
\[
E_{g}-\omega \approx \Delta _{g}. 
\]%
At the same time, the action of Eq.(\ref{actionme}) is of order one the
distance from the gap becomes of the order of the typical gap fluctuations 
\[
E_{g}-\omega \approx \delta E_{g}. 
\]%
It follows that the parameter that roughly determines which one of the two
mechanisms dominates is 
\begin{eqnarray}\label{beta0}
\beta =\frac{\Delta _{g}}{\delta E_{g}}. 
\end{eqnarray}
For $\beta \gg 1$ the asymptotic tails are dominated by mesoscopic
fluctuations and the result obtained in Ref.~[\onlinecite{Lamacraft}]
applies. In contrast, when $\beta \ll 1$ fluctuations of the impurity
concentration dominate the physics, and Eq.(\ref{finalzeroDOS}) describe the
asymptotic tails.

In more detail, if $\beta \ll 1$, as the energy $E$ is decreased from $E_{g}$
towards the Fermi level, beyond $E_{g}-E\approx \Delta _{g}$ the mesoscopic
gap fluctuations~become exponentially rare. However, for $\Delta
_{g}<E_{g}-E<\delta E_{g}$ the system is still in the range of the typical
gaussian fluctuations of $E_{g}$ associated to fluctuations of $\zeta $,
which dominate the physics. As the energy is decreased further, both
asymptotic result are applicable, and to determine which wins, one has to
compare the actions. The direct comparison shows that in the range $\delta
E_{g}\ll E_{g}-E\ll \delta E_{g}/\beta ^{3}$ the density of states is
dominated by the action associated to optimal fluctuations of $\zeta $. At
the crossover point $E_{g}-E\simeq \delta E_{g}/\beta ^{3}$, both actions
are of the same order $S\simeq S_{SUSY}\simeq 1/\beta ^{6}\gg 1$ but 
typically at this point the density of states is negligible. For example,
for $\beta =0.3$, at the crossover $S\simeq 10^{3}$, and $exp[-S]$ is
practically zero. Therefore, we conclude that the simple rule to determine
which mechanism dominates is to compare the typical gap fluctuations to the
effective level spacing: the largest wins.

\subsection{Bulk Optimal fluctuations.}

\label{sec2}

In this section, we study the effect of spatial fluctuations of the
spin-scattering rate in relation to subgap localized states. In particular,
we will consider a dirty superconductor containing both normal and
magnetic impurities, under the following conditions 
\begin{eqnarray}  \label{conditions}
l \ll \xi \;\;\;\;\;\;\zeta \ll 1\;\;\;\mathrm{and}\;\;\;n_{imp}\xi^d \gg 1,
\end{eqnarray}
where $l$ is the mean free paths relative to non-magnetic scattering, $\xi$
is the coherence length, $d$ is the dimensionality of the problem, and $%
n_{imp}$ is the average concentration of magnetic impurities in the system.
This set of conditions describe a disordered superconductor containing a
relatively large number of weak magnetic impurities.

The condition $l\ll \xi $ implies that the problem can be studied in the
framework of the semiclassical approximation~\cite{Kopnin}. In particular,
parametrizing both the semiclassical Green's function $g(r,\epsilon )$ and
anomalous Green's function $f(r,\epsilon )$ in terms of a phase $\theta $ by 
\begin{eqnarray}
g(\mathbf{r},\epsilon ) &=&\cos [\theta (\mathbf{r},\epsilon )], \\
f(\mathbf{r},\epsilon ) &=&i\sin [\theta (\mathbf{r},\epsilon )],
\end{eqnarray}%
one gets the Usadel equation \cite{Usadel,Kopnin} 
\begin{eqnarray} \label{Usadel}
&& \nabla ^{2}\theta (\mathbf{r})+i\epsilon \sin [\theta (\mathbf{r})]-\cos
[\theta (\mathbf{r})]-\nonumber \\ 
&&\zeta \sin [\theta (\mathbf{r})]\cos [\theta (\mathbf{r})] =0.  
\end{eqnarray}%
Here the unit length is $\xi =\sqrt{D/2\Delta }$, and $D$ is the diffusion
constant, and $\epsilon =E/\Delta $.

In this section, we will study small deviations from the solutions of the 
\textit{\ uniform} Usadel equation 
\begin{equation}
(i\epsilon -\frac{\zeta }{2}\cos (\theta ))\sin (\theta )-(1+\frac{\zeta }{2}%
\sin (\theta ))\cos (\theta )=0.  \label{uniformUsadel}
\end{equation}%
Let us derive a few well known properties of such solutions, corresponding
to the Abrikosov-Gorkov mean field theory~\cite{Abrikosov}. Setting $\theta
=-\pi /2-i\;\mathrm{arctanh}(u)$, we have 
\begin{eqnarray}
\sin (\theta ) &=&-\frac{1}{\sqrt{1-u^{2}}}, \\
\cos (\theta ) &=&-i\frac{u}{\sqrt{1-u^{2}}}.
\end{eqnarray}%
Therefore, the solution of Eq.~(\ref{uniformUsadel}) are all those $u$ such
that 
\[
\epsilon =u\left( 1-\zeta \frac{1}{\sqrt{1-u^{2}}}\right) . 
\]%
From this equation we obtain that the gap edge is at $\epsilon _{0}=(1-\zeta
^{2/3})^{3/2}$. In particular, the density of states per unit volume 
\[
\nu =\nu _{0}\mathrm{Re}[\cos (\theta )], 
\]%
is uniform and equal to zero for $\mid \epsilon \mid <\epsilon _{0}$. At the
gap edge the parameter $u$ is equal to $u_{0}=(1-\zeta ^{2/3})^{1/2}$.
Correspondingly, one has the solution $\theta _{edge}=-\pi /2-i~\;\mathrm{%
arctanh}(u_{0})$.

The Usadel equation is obtained neglecting large scale fluctuations of the
impurity concentration. This assumption limits the scattering to be
homogeneous in the appropriate long wavelength limit. On the other hand, in
a realistic system the distribution of impurities is poissonian. This
implies that the number of impurities in any finite volume element
fluctuates, with a variance equal to the average.

Following this observation, we introduce an effective theory describing the
fluctuations of the concentration of magnetic impurities on length scales of
the order, or larger, than the coherence length. In terms of the Usadel
equation Eq.(\ref{Usadel}) $\zeta $ becomes a position dependent random
variable. As a consequence of the Poissonian statistics of the impurities,
we have 
\begin{eqnarray}
\zeta (\mathbf{r}) &=&\zeta +\delta \zeta (\mathbf{r}), \\
\langle \delta \zeta \rangle &=&0, \\
\langle \delta \zeta (\mathbf{r})\delta \zeta (\mathbf{r^{\prime }})\rangle
&=&\left( \frac{\zeta ^{2}}{n_{imp}\;\xi ^{d}}\right) \;\delta (\mathbf{r}-%
\mathbf{r^{\prime }}),
\end{eqnarray}%
where $\zeta $ is the average, uniform, dimensionless, spin-scattering rate,
and $d$ is the dimensionality.

Let us look for solutions of Eq.(\ref{Usadel}) at energies slightly below
the gap ($\epsilon =\epsilon _{0}-\delta \epsilon $) in the form 
\[
\theta (\mathbf{r})=\theta _{edge}-i\phi (\mathbf{r}). 
\]%
Expanding Eq.(\ref{Usadel}) in $\phi ,\delta \epsilon $, and $\delta \zeta $%
, in the limit $\zeta \ll 1$ one obtains 
\begin{eqnarray}\label{expUsadel1}
\nabla ^{2}\phi +\frac{3}{2}\zeta ^{1/3}\phi ^{2}=\frac{1}{\zeta ^{1/3}}%
\left( \delta \epsilon -\frac{1}{\zeta ^{1/3}}\delta \zeta \right) . 
\end{eqnarray}
Rescaling again the length in units $\lambda =\xi \;(2/3)^{1/4}$, and
defining $\psi =(3/2\;\;\zeta ^{2/3})^{1/2}\phi $, we recast Eq.~(\ref%
{expUsadel1}) in the simpler form 
\begin{eqnarray}\label{expUsadel}
\nabla ^{2}\psi +\psi ^{2}=\delta \epsilon -f(\mathbf{r}), 
\end{eqnarray}
where 
\[
f(\mathbf{r})=\frac{\delta \zeta }{\zeta ^{1/3}}. 
\]%
In particular, 
\begin{eqnarray}
\langle f(\mathbf{r})f(\mathbf{r^{\prime }})\rangle &=&\eta \;\delta (%
\mathbf{r}-\mathbf{r^{\prime }}),  \nonumber \\
\eta &\equiv &\left( \frac{\zeta ^{4/3}}{n_{imp}\;\xi ^{d}}\left( \frac{3}{2}%
\right) ^{d/4}\right) .  \label{eta}
\end{eqnarray}

Let us split $\psi $ as $\psi =-x+iy$. Then we have the system 
\begin{eqnarray}
-\nabla ^{2}x+x^{2}-y^{2} &=&\delta \epsilon -f(\mathbf{r}),
\label{saddlepoint1} \\
-\frac{1}{2}\nabla ^{2}y+x\;y &=&0.  \label{saddlepoint2}
\end{eqnarray}%
Interestingly, this set of equations is analogous to the equations obtained
by Larkin and Ovchinikov~\cite{Larkin}, in the context of the study of gap
smearing in inhomogeneous superconductors.

Notice that the analytical properties of the quasi-classical Green's
functions impose some constraints on the solutions of these equations.
Indeed, the DOS is 
\begin{eqnarray}  \label{DOS1}
\nu(\mathbf{r})/\nu_0=\mathrm{Re }[\cos(\theta)] \simeq \sqrt{\frac{2}{%
3~\zeta^{4/3}}}\;y(\mathbf{r}),
\end{eqnarray}
where $\nu_0$ is the bare DOS per unit volume at the Fermi level. Therefore,
we must have $y(\mathbf{r})>0$.

Our aim is to evaluate the average DOS $\langle \nu (\mathbf{r})\rangle /\nu
_{0}=\sqrt{2/3~\zeta ^{4/3}}\langle y(\mathbf{r})\rangle $ at a distance $%
\delta \epsilon $ below the average gap. In particular, 
\begin{equation}
\langle y(\mathbf{r})\rangle =\frac{\int \;\mathcal{D}[f(\mathbf{r})]\;y(%
\mathbf{r},\delta \epsilon \mid f(\mathbf{r}))\;e^{-\frac{1}{2\eta }\;\int d%
\mathbf{r}(f(\mathbf{r}))^{2}}}{\int \;\mathcal{D}[f(\mathbf{r})]\;e^{-\frac{%
1}{2\eta }\;\int d\mathbf{r}(f(\mathbf{r}))^{2}}}.  \label{<y(r)>}
\end{equation}

For energies below, but not too close, to the average gap edge, the leading
contribution to the DOS comes from exponentially rare fluctuations of $f(%
\mathbf{r})$, describing a local increase of the impurity concentration, and
therefore a local suppression of the gap below its average value. In
particular, we have to find the \textit{optimal} fluctuation of $f(\mathbf{r}%
)$, i.e. the configuration of $f(\mathbf{r})$ associated to the dominant
contribution to $\langle \nu(\mathbf{r}) \rangle$.

In order to be characterized by a finite action, an optimal fluctuation
has to be such that $f(\mathbf{r})\rightarrow 0$ as $r=~\mid ~\mathbf{r}%
~\mid \rightarrow \infty $. The asymptotic behavior of any solution of Eqs.(%
\ref{saddlepoint1}-\ref{saddlepoint2}) with $f(r)$ describing an optimal
fluctuation is therefore 
\begin{eqnarray}
y(r) &\rightarrow &0\;\;\;\mathrm{as}\;\;\;r\rightarrow +\infty ,
\label{BC1} \\
x(r) &\rightarrow &\sqrt{\delta \epsilon }\;\;\;\mathrm{as}%
\;\;\;r\rightarrow +\infty .  \label{BC2}
\end{eqnarray}%
Using Eq.(\ref{saddlepoint2}), the second condition implies 
\begin{equation}
\frac{\nabla ^{2}y}{2y}\rightarrow \sqrt{\delta \epsilon }\;\;\;\mathrm{as}%
\;\;\;r\rightarrow +\infty .  \label{BCy}
\end{equation}%
Since the system is diffusive, and scattering is isotropic, it is natural to
assume a spherically symmetric optimal fluctuation, its scale being $\lambda 
$, i.e. $f(r)=f~(~r~/~\lambda ~)$. Then $y(r)=y(r/\lambda )$ which implies 
\[
\frac{\nabla _{r}^{2}y}{2y}=1/\lambda ^{2}\frac{\nabla _{(r/\lambda )}^{2}y}{%
2y}. 
\]%
Thus Eq.(\ref{BCy}) implies $\lambda \propto (\delta \epsilon )^{-1/4}$.

The problem above has many analogies to the problem of localized states in
the Lifshits tails of a disordered system~\cite{Halperin,Lax,Zittarz} [see
Appendix A]. The analogy becomes clear if we substitute $V(r)=x-\sqrt{\delta
\epsilon }$ and rewrite Eq.(\ref{saddlepoint2}) as 
\begin{eqnarray}
\left[ -\frac{1}{2}\nabla ^{2}+V(r)\right] y(r) &=&-\sqrt{\delta \epsilon }%
\;y(r),  \label{Sch1} \\
V(r) &\rightarrow &0\;\;\;\;\mathrm{as}\;\;\;\;\;r\rightarrow +\infty . 
\nonumber
\end{eqnarray}%
As in the Lifshits tails problem, we are essentially looking for a potential
well $V(r)$ that admits a ground state at energy $-\sqrt{\delta \epsilon }$.
A ground state condition is important because the nodes in the "wave
function" $y(r)$ are not consistent with the analytical properties of the
quasi-classical Green's functions. Despite the close analogy, it is
important to notice two differences with the Lifshits tails problem: \emph{%
(i)} our wave function $y(r)$ is a component of the Usadel phase directly
proportional to the DOS, and \emph{\ (ii)} the statistical weight of the
potential in the standard Lifshits tails problem is $S\propto \int
dr\;(V(r))^{2}$ while here it is 
\begin{eqnarray}
&&  \nonumber \\
S &=&\frac{1}{2\eta }\int d\mathbf{r}\;(f(r))^{2}  \nonumber \\
&=&\int \frac{1}{2\eta }d\mathbf{r}\;(y^{2}+\nabla ^{2}V-V^{2}-2\sqrt{\delta
\epsilon }V)^{2},  \label{S}
\end{eqnarray}%
where we used Eq.(\ref{saddlepoint1}) to express $f(r)$ in terms of $V(r)$
and $y$.

With the exponential accuracy one can neglect the $y^{2}$ term in Eq.(\ref{S}%
) because the integration over the fluctuations near the saddle point
characterized by Eq.(\ref{Sch1}) leads to a non-zero contribution to the
density of states Eq.(\ref{<y(r)>}) [see Appendix B]. To find the parametric
dependence of the action, we note that the Schr\"{o}dinger equation Eq.(\ref{Sch1})
implies that the optimal fluctuation has a scale $\lambda \propto (\delta
\epsilon )^{-1/4}$ and an amplitude $V_{0}(r)\propto -\sqrt{\delta \epsilon }$ .
Writing $V=\sqrt{\delta \epsilon }\;v(r/\lambda )$ we get 
\begin{eqnarray}
S &\simeq &\frac{1}{2\eta }\int d\mathbf{r}\left( \nabla ^{2}V-V^{2}-2\sqrt{%
\delta \epsilon }V\right) ^{2}  \nonumber \\
&=&\frac{1}{2\eta }(\delta \epsilon )^{2-d/4}\int d(\mathbf{r}/\lambda
)\left( \nabla ^{2}v-v^{2}-2v\right) ^{2}  \nonumber \\
&=&\mathrm{const}\frac{n_{imp}\xi ^{d}}{\zeta ^{4/3}}(\delta \epsilon
)^{2-d/4},  \label{S_scale}
\end{eqnarray}%
where $\eta $ is given by Eq.(\ref{eta}).

We now find the exact shape of the optimal fluctuation and the resulting
numerical constant in Eq.(\ref{S_scale}) from the Euler-Langrange equations. We
have to look for a the most probable $f(r)$ which when inserted in Eq.(\ref%
{saddlepoint1}-\ref{saddlepoint2}), produces a solution $y(r)$, positive
everywhere, and such that $\nabla ^{2}y/(2y)\rightarrow \sqrt{\delta
\epsilon }$ at infinity. This general strategy can be applied to the
classical problem of Lifshits tails, as explained in Appendix.A.

Expressing $f(r)$ in terms of $x$ and $y$, we get the dimensionless action 
\begin{eqnarray}
S &=&\frac{1}{2\eta }\int d\mathbf{r}\left[ \nabla ^{2}x-x^{2}+\delta
\epsilon \right] ^{2},  \label{actionmin} \\
x &=&\frac{\nabla ^{2}y}{2y},
\end{eqnarray}%
where $d$ is the dimensionality. This action admits a trivial stationary
point corresponding to zero action or $f(r)=0$ in Eqs (\ref{saddlepoint1}-%
\ref{saddlepoint2}) characterized by%
\begin{equation}
x^{\prime \prime}+\frac{d-1}{r}\;x^{\prime}-x^{2}+\delta \epsilon =0,
\label{trivialsaddle}
\end{equation}%
where the prime indicates the derivative with respect to $r$.
It is possible to show that this equation admits  instanton solutions.
However, these solutions have either $y(r)=0$ at every point in space, or $%
y(r)<0$ somewhere. The analytic properties of quasi-classical Green's
functions indicate that the latter solutions are not allowed. 
On the other hand, the solutions with $y(r)=0$ do not contribute 
to the DOS in the present approximation scheme. However, 
these solutions become meaningful in the
Sigma model approach to this problem; there they describe the saddle point
corresponding to mesoscopic fluctuations generating subgap states~\cite%
{Lamacraft, Ostrovsky, Dima}.

Let us now find the equations associated with the nontrivial saddle points.
In spherical coordinates, the Euler-Lagrange equation corresponding to the
action, Eq.(\ref{actionmin}), is 
\begin{eqnarray}
&&\partial _{r}^{2}\left( r^{d-1}\;\mathcal{A}\right) -(d-1)\;\partial
_{r}\left( r^{d-2}\;\mathcal{A}\right)  \nonumber \\
&&-2\;x\;r^{d-1}\mathcal{A}=0,  \label{NonTrivialSaddle} \\
&&\mathcal{A}[x,x^{\prime },x^{\prime \prime },r]=x^{\prime \prime }+\frac{%
d-1}{r}x^{\prime }-x^{2}+\delta \epsilon .
\end{eqnarray}%
The problem of finding the instanton solutions of this equation describing
the optimal fluctuations is considerably simplified by the observation that,
in all dimensions [$d=1,2,3$], the solutions of 
\begin{equation}
\mathcal{A}[x,x^{\prime },x^{\prime \prime },r]=\frac{2\;x^{\prime }}{r},
\label{Ansatz}
\end{equation}%
are also solutions of Eq.(\ref{NonTrivialSaddle}). Therefore, the nontrivial saddle
points~\cite{foot} are described by the system 
\begin{eqnarray}
x^{\prime \prime }+\frac{d-3}{r}\;x^{\prime }-x^{2}+\delta \epsilon &=&0,
\label{EL1} \\
y^{\prime \prime }+\frac{d-1}{r}\;y^{\prime } &=&x\;y.  \label{EL2}
\end{eqnarray}%
It is interesting to notice that the equation above for the variable $x(r)$
coincides with the equation describing the trivial saddle point, Eq.(\ref%
{trivialsaddle}), in $d-2$ dimensions. In three dimensions we get 
\begin{eqnarray}
x_{0} &=&\sqrt{\delta \epsilon }\left( 1-3\;(\mathrm{sech}\left[ r/\lambda %
\right] )^{2}\right) , \\
y_{0} &\propto &\;\frac{1}{r/\lambda }\tanh \left[ r/\lambda \right] (%
\mathrm{sech}\left[ r/\lambda \right] )^{2}, \\
\lambda &=&\left( \frac{\delta \epsilon }{4}\right) ^{-1/4}.
\end{eqnarray}%
The optimal fluctuation of the dimensionless spin-flip scattering rate can
be then evaluated using 
\[
f_{0}(r)=\delta \epsilon -\left( \frac{\nabla ^{2}y_{0}}{2y_{0}}\right)
^{2}+\nabla ^{2}\left( \frac{\nabla ^{2}y_{0}}{2y_{0}}\right) 
\]%
Thus, in three dimensions, the action at the nontrivial saddle point is 
\begin{eqnarray}
S &=&\frac{4\pi }{2\eta }\left( \frac{\delta \epsilon }{4}\right) ^{5/4}%
\frac{384}{5}  \nonumber  \label{3d} \\
&\simeq &63\left[ \frac{n_{imp}\;\xi ^{3}}{\zeta ^{4/3}}\right] \delta
\epsilon ^{5/4}.
\end{eqnarray}

In one and two dimensions the solution of Eq.(\ref{EL1}-\ref{EL2}) is not so
straightforward. In order to estimate the action of the optimal fluctuation,
we had to minimize the action $S$ in terms of a variational ansatz $%
y_{a,b}(r)$ depending on two free parameters. The asymptotic form of any
variational ansatz $y_{a,b}(r)$ must be 
\begin{eqnarray}
y(r) &\rightarrow &\frac{e^{2r/\lambda }}{\sqrt{r/\lambda }}\;\;\;\;\;\;(d=2)
\\
y(r) &\rightarrow &e^{2r/\lambda }\;\;\;\;\;\;(d=1)
\end{eqnarray}%
in order to obtain to a finite action. The result of this calculation is 
\begin{eqnarray}
S &\simeq &14\left[ \frac{n_{imp}\;\xi ^{2}}{\zeta ^{4/3}}\right] \delta
\epsilon ^{3/2}\;\;\;\;(d=2)  \label{2d} \\
S &\simeq &2.8\left[ \frac{n_{imp}\;\xi }{\zeta ^{4/3}}\right] \delta
\epsilon ^{7/4}\;\;\;\;(d=1)  \label{1d}
\end{eqnarray}

Clearly $\langle y(r)\rangle \propto \exp [-S]$. This equality must be
supplemented by the calculation of the prefactor, i.e. gaussian
fluctuations, using the standard technique due to Zittarz and Langer \cite%
{Zittarz}. This calculation is reported in some detail in Appendix.B. The
result is 
\begin{equation}
\frac{\langle \nu \rangle }{\nu _{0}}\simeq \tilde{b}_{d}\frac{\sqrt{%
n_{imp}\xi ^{d}}}{\zeta ^{4/3}}\delta \epsilon ^{\alpha _{d}}\exp \left[ -%
\tilde{a}_{d}\frac{n_{imp}\xi ^{d}}{\zeta ^{4/3}}\delta \epsilon ^{2-d/4}%
\right]  \label{finalDOS}
\end{equation}%
where the exponent is $\alpha _{d}=1/8(d(10-d)-12)$. The values of the
numerical coefficient of the action $\tilde{a}_{d}$, as well as the
estimated value of the numerical constants $\tilde{b}_{d}$ according to the
calculation reported in Appendix.B, are summarized in the following table

\begin{eqnarray}
\begin{array}{|c||c|c|}
\hline
\mathrm{dimensions} & \;\;\;\;\;\;\;\tilde{a}_d\;\;\;\;\;\;\; & 
\;\;\;\;\;\;\;\tilde{b}_d\;\;\;\;\;\;\; \\ \hline\hline
1 & 2.8 & 0.0007 \\ \hline
2 & 14 & 0.006 \\ \hline
3 & 63 & 0.2 \\ \hline
\end{array}
\nonumber
\end{eqnarray}

First of all, the result of Eq.(\ref{finalDOS}) is asymptotic, and requires $%
S \gg 1$. This condition translates into $\delta\epsilon \gg
\delta\epsilon_0 $, where 
\begin{eqnarray}  \label{limit}
\delta\epsilon_d \equiv \left[\frac{1}{\tilde{a}_{d}} \frac{\zeta^{4/3}}{
n_{imp}\xi^d} \right] ^{\frac{4}{8-d}}.
\end{eqnarray}
Since we assumed $\zeta \ll 1$, $n_{imp}\xi^d \gg 1$ , one has $%
\delta\epsilon_0 \ll 1$.

As expected, the final result given by (\ref{finalDOS}) crosses over to the
result of the zero dimensional calculation performed in Sec.~\ref{sec1} in
the limit $d\rightarrow 0$. The relation between the $d\neq 0$ and $d=0$
results becomes simpler when we rewrite Eq.(\ref{finalDOS}) as 
\begin{equation}
\frac{\langle \nu \rangle }{\nu _{0}}\propto \exp \left[ -\tilde{a}%
_{d}\left( \frac{\lambda _{0}}{L}\right) ^{d}\left( \frac{E_{g}-E}{\delta
E_{g}}\right) ^{2}\right] ,  \label{compactDOS}
\end{equation}%
where $\lambda _{0}=\xi /(\delta \epsilon )^{1/4}$ [see Eq.(\ref{size})].
Though this way of writing the result makes the $d\rightarrow 0$ limit
transparent, it should be kept in mind that the action in finite dimensional
systems does not depend on the sample size $L$ but only on intensive
quantities.

Let us now compare the asymptotic tails resulting from fluctuations of $%
\zeta $, Eq.(\ref{finalDOS}) with those associated to mesoscopic gap
fluctuations~\cite{Lamacraft}. The latter give an asymptotic DOS 
\begin{eqnarray}
\frac{\langle \nu \rangle }{\nu _{0}} &\propto &\exp \left[ -S_{SUSY}\right]
,  \nonumber  \label{finalDOSSUSY} \\
S_{SUSY} &=&a_{d}\left( \frac{4\pi \nu _{0}\Delta \xi ^{d}}{\zeta ^{2/3}}%
\right) (\delta \epsilon )^{3/2-d/4},
\end{eqnarray}%
where $a_{d}$ is a numerical constant. In one dimension it is given by $%
a_{1}\approx 8^{4}\sqrt{24}/5\approx 4000$. Introducing $\lambda _{0}=\xi
/(\delta \epsilon )^{1/4}$, one may write 
\[
S_{SUSY}\simeq a_{d}\left( \frac{\lambda _{0}}{L}\right) ^{d}\left( \frac{%
E_{g}-E}{\Delta _{g}}\right) ^{3/2}, 
\]%
with $\Delta _{g}$ given by Eq.(\ref{levspacing}). From this equation we
conclude that the ratio between this action and the action of the optimal
fluctuations of $\zeta $ [Eq.(\ref{compactDOS})] is independent on
dimensionality (apart from a numerical prefactor). This is a direct
consequence of the fact that the typical linear sizes of the optimal
fluctuations associated to the two mechanisms are identical.

In order to establish which fluctuation dominates the physics of the subgap
tails, we first notice that $S_{SUSY}$ is of order of unity at $\delta
\epsilon \simeq \delta \epsilon _{d}^{\prime }$, where 
\[
\delta \epsilon _{d}^{\prime }\equiv \left( \frac{1}{a_{d}}\frac{\zeta ^{2/3}%
}{4\pi \nu _{0}\Delta \xi ^{d}}\right) ^{\frac{4}{6-d}}. 
\]%
Therefore, in a finite dimensional system, the parameter that determines
which one of the two mechanisms dominates is 
\begin{equation}
\beta _{d}\equiv \frac{\delta \epsilon _{d}^{\prime }}{\delta \epsilon _{d}}.
\label{betad}
\end{equation}

As in the zero dimensional case, when $\beta _{d}\gg 1$ the subgap tails are
dominated by mesoscopic gap fluctuations, and the asymptotics of the subgap
tails is described by Eq.(\ref{finalDOSSUSY}). In contrast, when $\beta
_{d}\ll 1$ the physics is dominated by the fluctuations of the concentration
of magnetic impurities, and Eq.(\ref{finalDOS}) applies. More precisely, as
the energy is lowered from the gap edge, i.e. as $\delta \epsilon $
increases, first the asymptotic result relative to mesoscopic gap
fluctuations starts being applicable beyond $\delta \epsilon \simeq \delta
\epsilon _{d}^{\prime }$. However, for $\delta \epsilon _{d}^{\prime
}<\delta \epsilon <\delta \epsilon _{d}$ the system is within the range of
typical fluctuations of $\zeta $, which dominate over the exponential tails
associated to mesoscopic gap fluctuations. Increasing $\delta \epsilon $
beyond $\delta \epsilon _{d}$ both asymptotic results are applicable and the
two actions should be compared. In particular, 
\[
\frac{S}{S_{SUSY}}=\frac{(\delta \epsilon _{d}^{\prime })^{\frac{6-d}{4}}}{%
(\delta \epsilon _{d})^{\frac{8-d}{4}}}\;\delta \epsilon ^{1/2}, 
\]%
implying that for 
\[
\delta \epsilon _{d}\ll \delta \epsilon \ll \delta \epsilon \left( \frac{1}{%
\beta _{d}}\right) ^{\frac{6-d}{2}}, 
\]%
the asymptotic tails are dominated by fluctuations of $\zeta $. At the
crossover, i.e. $\delta \epsilon \simeq (\delta \epsilon _{d})^{\frac{8-d}{2}%
}/(\delta \epsilon _{d}^{\prime })^{\frac{6-d}{2}}$, both actions are 
\[
S\simeq S_{SUSY}\simeq \left( \frac{1}{\beta _{d}}\right) ^{\frac{(6-d)(8-d)%
}{8}}>>>1. 
\]%
This implies that, at the crossover, the density of states is already
negligible. For example, for $d=2$ and at the crossover point, $S\simeq
(1/\beta _{d})^{3}$. Taking, $\beta _{d}=0.1$, one obtains at the crossover $%
S\approx 10^{3}$, implying that $\exp (-S)$ is practically zero.

\section{Shallow quasiparticle traps in superconducting Qubits.}

~\label{sec2b}

In the previous sections we found that a low concentration of the
paramagnetic impurities lead to a small but finite density of the localized
subgap quasiparticle states. In this section, we discuss the effect of these
states on small superconducting devices, especially the ones proposed
for quantum computation. Because the density of these localized states is
always very low, they do not affect much the macroscopic properties such as
specific heat or Josephson current of big junctions. However, even a small
number of quasiparticles trapped in these states might affect the long time
phase coherence in the Cooper pair boxes \cite{Nakamura},\cite%
{Devoret,Martinis2} or flux qubits\cite{Mooji}, so we shall focus on such
devices here.

\begin{figure}[tbp]
\resizebox{7.5 cm}{!}{\includegraphics{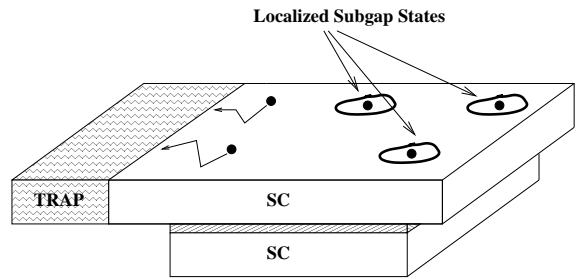}}
\caption{ A pictorial representation of a Josephson junction
composed of two superconducting electrodes separated by an oxide layer, the
basic building block of superconducting Qubits. In these system,
quasiparticles are typically excited in the Qubit duty cycle. Delocalized
quasiparticle diffuse through the system towards the metallic quasiparticle
traps. In turn, quasiparticles trapped in localized subgap states remain
within the system. During subsequent Qubit operations, these localized
quasiparticles represent active degrees of freedom that may contribute to
decoherence. Our estimates indicate that, for a junction with aluminum
electrodes of size $\simeq 10 \times 10 \times 0.1 \;{\ \protect\mu m}^3$
containing $\approx 1\;{\ ppm}$ of magnetic impurities, the average number
of localized subgap states is about $10^2-10^3$ per electrode. 
The phonon-limited lifetime of a
quasiparticle in such states is expected to be $\approx 100\;s$ at $T=10\;%
{\ mK}$. On the other hand, we estimate the presence of
$10^2-10^3$ localized quasiparticles coupled to phase fluctuations in the
junction to be limit the Qubit quality factor to $10^4-10^5$,
for an operational frequency of $10\;{\rm GHz}$.}
\label{fig:fig1}
\end{figure}

The building blocks of these circuits are typically $\mathrm{%
Al-Al_{2}O_{3}-Al}$ or $\mathrm{Nb-Al_{2}O_{3}-Nb}$ Josephson junctions.
When quasiparticles are generated, say, during the readout process, most of
them quickly recombine and disappear in the condensate. In order to
eliminate the remaining quasiparticles, superconducting circuits incorporate
normal metal leads acting as quasiparticle traps ~\cite{Devoret, Martinis1}.
At low temperatures, these leads can be viewed as the sinks for all the
quasiparticles that may diffuse to them. However, a quasiparticle localized
within a superconducting electrode cannot diffuse to the quasiparticle
traps; instead it remains localized inside the superconducting circuit and
represents an active degree of freedom that may contribute to decoherence in
subsequent operations of the qubit [see Fig.~\ref{fig:fig1}].

We begin with the estimate of the number of localized states in a typical
experimental setup. We focus on aluminum based qubits. The typical
parameters for aluminum are $\Delta =200\;\mathrm{\mu eV}$, $D=50\;\mathrm{%
cm^{2}/s}$, $v_{F}=2 \times 10^{8}\;\mathrm{cm/s}$, and $\nu _{0}\approx
1.5 \times 10^{22}\;\mathrm{cm^{-3}\;eV^{-1}}$. This implies that the coherence
length is $\xi \approx 0.1\;\mathrm{\mu m}$, while the mean free path is $%
l\approx 0.01\;\mathrm{\mu m}\ll \xi $. Therefore, the system is diffusive,
and the first condition for the applicability of the results obtained in
previous sections is satisfied [see Eq.(\ref{conditions})]. Further, because
the thickness of the samples is about or less than the coherence length,
they are effectively two dimensional.

The strength of the spin-flip scattering and the number of paramagnetic
impurities is more difficult to estimate reliably. The problem is that for a
typical sample of thickness $l\approx 0.1\mu m$  surface scattering is
rather important and it is very likely that there it contains a significant
spin-flip contribution. Indeed, the surface of the $Al$ devices is covered
by a thin layer of glassy $Al_{2}O_{3}$. A typical glass has a concentration
of $n_{TLS}\sim 10^{7}-10^{8}\mu m^{-3}$ of two level systems which are
usually attributed to the trapped electrons \cite{Burin1995}. This
translates into a surface density of $10^{3}-10^{4}\mu m^{-2}$ of free
spins in the boundary layer of $Al_{2}O_{3}$ that interact with the
electrons in the metal or into the effective density $10^{4}-10^{5}\mu
m^{-3} $ \ per unit volume of the device. Because this estimate counts only
the spins associated with the two level systems in a glass, we expect that
the actual number is somewhat bigger. On the other hand, assuming that the
strength of the spin-spin interaction $J\approx 0.1\;eV$ and its range $%
r\approx 4\;\mathring{A}$ we get that the volume density $n_{sf}$\textbf{$%
\simeq $}$10^{5}\mu m^{-3}$ leads to a spin-scattering time of \textbf{$\tau
_{s}\approx $}$10^{-8}\;s~$\ which is the lower bound given by the weak
localization magnetoresistance data \cite{Gershenson}. Thus,
we believe that the realistic estimate for the effective density of the
paramagnetic impurities in the wires of the $0.1\mu m$ thickness is about $%
n_{sf}$\textbf{$\simeq $}$10^{5}\mu m^{-3}$.

For this value of the the paramater $\beta $ [see Eq. (\ref{betad})]
is very small: $\beta _{2}\approx 10^{-5}\ll 1$. Therefore, the asymptotics
described by Eq.(\ref{finalDOS}) applies. Using now Eq.(\ref%
{finalDOS}) for the density of states we estimate the total number of the
localized states per unit area in these conditions: $\rho \sim 1-10\;\mu
m^{-3}$. This means that for the large junctions of the area $A\approx
10\times 10\;\mu m^{2}$, one gets about $N_{sub}\approx 10^{2}-10^{3}$
localized states on each superconducting electrode. The length scale
associated to these states is about $1\mu m$. With the parameters above, we
expect these localized states to be very shallow, typically $0.01\;\mu eV$
below the gap edge. This energy scale is two orders of magnitude smaller
than the typical temperature in qubit experiments, $T\simeq 10\;mK\simeq
1\;\mu eV/k_{B}$. This implies that if a non-equilibrium quasiparticle
generated in the duty cycle of the qubit gets trapped into one of these
localized states, it will eventually absorb a phonon, get excited above the
mobility edge, and diffuse out of the sample. However, at such low
temperatures, the time scale $\tau _{tr}$ associated to these
phonon-assisted detrapping processes is very long, $\tau _{tr}\approx
100s\;\;$~\cite{tau,Lawrence} . At the same time, in a superconducting qubit
composed of large junctions of area $10\times 10\;\mu m^{2}$ and operating
at a frequency $f=10\;GHz$ the qubit relaxation rate associated to the
coupling between the phase degrees of freedom and a single localized
quasiparticle is $\Gamma _{qp} \approx 10^{3}\;Hz$ [see Appendix C]. 
Therefore, the presence of $10^{2}-10^{3}$ localized quasiparticles interacting with the
phase degrees of freedom across the junction is expected to limit the
quality factor to be $Q=\nu _{0}/\Gamma _{qp}\approx 10^{4}-10^{5}$ . It is
important to mention that for smaller systems, e.g. of area $A\approx
1\times 1\;\mu m^{2}$, the probability of having a single localized subgap
state of area considerably smaller than $A$ is negligible. The subgap tails
are effectively zero dimensional and the tail states extend to the whole
system. So, an electron trapped in such state can always escape to the lead
through the Josephson contact.

\section{Conclusions.}

~\label{sec3}

We have shown that in a diffusive superconductor containing magnetic
impurities at small concentrations [$\zeta <1$], the subgap tails of the DOS
have two distinct contributions, one due to the universal mesoscopic gap
fluctuations~\cite{Lamacraft}, the second resulting from inhomogeneous
fluctuations of the spin scattering time $\zeta $.

We calculated the contribution to the density of states due to the
fluctuations of $\zeta $ [see Eq.(\ref{finalzeroDOS}) for $d=0$ and Eq.(\ref%
{finalDOS}) for $d>0$] and established, through a direct comparison of the
results, the parameter that controls which of the two mechanisms dominates
the physics of subgap states [see Eq.(\ref{beta0})-(\ref{betad})].
The two mechanism are related by dimensional reduction, both at the 
level of instanton equations and in the dependence of the actions on $\delta\epsilon$. 
The deep reason behind the occurrence of dimensional reduction in this 
context is presently unclear, and requires further study. 

On the theoretical side, we believe that the mechanism generating subgap
localized states studied in this work is not limited to the dirty
superconductors containing paramagnetic impurities. For example, our study
is clearly related to the study of inhomogeneous superconductors performed
in a seminal paper by Larkin and Ovchinikov, Ref.[\onlinecite{Larkin}]. In
particular, in three dimensions they obtained result similar to (\ref%
{finalDOS}). It is interesting to notice that a recent solution of the same
problem by the replica Sigma model approach~\cite{Meyer} gave the same
asymptotic behavior as the mesoscopic gap fluctuations [Eq.(\ref{Lama})].
The source of this discrepancy was previously attributed to the
impossibility to construct in the present context a Lifshits argument~\cite%
{Meyer} that was used in the estimate of the action of the optimal
fluctuation in Ref.[\onlinecite{Larkin}]. In contrast, we have argued that
both results correctly describe two different physical effects that work in
parallel: mesoscopic Random-Matrix like gap fluctuations~\cite%
{Meyer} and long wavelength fluctuations of the coarse grained gap~\cite%
{Larkin}. In analogy with the present study, a direct comparison of the two
resulting actions determines which mechanism dominates the asymptotics of
the density of states.

Finally, we applied the theoretical results summarized above to estimate the
effects of a weak concentration of magnetic impurities [$\approx 1\;\mathrm{%
ppm}$, corresponding to $\zeta \approx 10^{-4}$] in Aluminum-based Qubits
and Josephson-junction arrays [see Fig.(\ref{fig:fig1})]. 
Our estimates indicate
that the presence of a small concentration of magnetic impurities is mostly
relevant for large qubits, designed around the $10\;{\mu m}$ scale. Under
the conditions above, Cooper pair boxes of the size $10\times 10\times 0.1\;%
\mathrm{\mu m^{3}}$ are expected to have an average of $\approx 10^{2}-10^{3}
$ two dimensional subgap localized states per electrode. At a base
temperature of $10\;\mathrm{mK}$, the lifetime of a non-equilibrium
quasiparticle localized in one of such states is expected to be as long as $%
100s$. We estimate that this would limit that quality factor of the qubit
operating at the frequency $f=10Ghz$ by $Q\approx 10^{4}-10^{5}$.

\section{Acknowledgments.}

We would like to thank also N. Andrei, M. Feigel'man, Y. Gefen, A.
Lamacraft, A. Schiller, and especially B. Altshuler, M. Gershenson, E.
Lebanon and M. M\"{u}ller for discussions. This work is supported by NSF
grant DMR 0210575.

\appendix

\section{Analogies to the problem of band tails in disordered conductors.}

In the following, we illustrate how the variational reasoning employed in
the main text works in the context of the problem of Lifshits tails, and how
the exact solution in one dimension for the leading exponential dependence
of the DOS tails relative to this problem is recovered~\cite%
{Halperin,Lax,Zittarz}.

Consider the Schr\"{o}dinger equation 
\begin{eqnarray}  \label{SchLif}
\left[-\frac{\nabla^2}{2}+V(r)\right]\;y(r)=-E\;y(r),
\end{eqnarray}
where $V(r)$ is a white noise potential 
\begin{eqnarray}
\langle V(r)V(r^{\prime}) \rangle=\gamma\;\delta(r-r^{\prime}).
\end{eqnarray}
Well below the band edge ($E=0$), the leading contribution to the DOS
associated to Eq.(\ref{SchLif}) comes from rare local fluctuations of $V(r)$
able to generate bound states. In order to find the leading exponential
dependence of the average DOS as a function of energy, we have to find the
most probable potential $V(r \mid E)$ admitting a bound state at energy $-E$%
, i.e the optimal fluctuation. The average DOS is then 
\begin{eqnarray}
\langle \rho \rangle &\propto& e^{-S},  \nonumber \\
S &=& \frac{1}{2\gamma}\int \;dr\;(V(r \mid E))^2.
\end{eqnarray}

Let us now look for a ground state at energy $-E$, i.e. assume that the wave
function $y(r)$ does not have nodes [and can therefore be chosen to be
positive]. For every given $y(r)$, we can use the Schr\"{o}dinger equation
to obtain 
\begin{eqnarray}  \label{Sch4}
V(r)=\frac{\nabla^2 y}{2y}-E.
\end{eqnarray}
This equation specifies for every positive, non-vanishing, and smooth $y(r)$%
, the potential well $V(r)$ that admits it as a ground state of energy $-E$.
Notice that $y(r)$ must tend to $0$ as $\mid r \mid \rightarrow +\infty$.
The same is true for $V(r)$, otherwise its action would be infinite. Thus 
\begin{eqnarray}
\frac{\nabla^2 y}{2y}\rightarrow E\;\;\;\mathrm{as}\;\;\;\mid r
\mid\rightarrow +\infty.
\end{eqnarray}

Consider now the functional 
\begin{eqnarray}
I[V(r)]=\int \;dr\;(V(r))^2.
\end{eqnarray}
Let use Eq.(\ref{Sch4}) to rewrite this functional as a functional of $y(r)$%
. One obtains 
\begin{eqnarray}  \label{cipolla}
I[y(r)]=\int \;dr\;\left(\frac{\nabla^2 y}{2y}-E\right)^2.
\end{eqnarray}
The problem of finding the most probable $V(r)$ admitting a ground state at
energy $-E$ is now reduce to the problem of finding the stationary points of 
$I[y(r)]$. Since the argument of the integral is squared, one might be
tempted to find the most obvious stationary point by setting 
\begin{eqnarray}  \label{ippo2}
\frac{\nabla^2 y}{2y}-E=0.
\end{eqnarray}
This procedure is obviously incorrect, since it corresponds to the search a
bound state in a flat potential landscape $V(r)=0$. In mathematical terms,
Eq.(\ref{ippo2}) does not admit nontrivial solutions satisfying the
appropriate boundary conditions introduced above.

Therefore, let us go back to Eq.(\ref{cipolla}) and find the nontrivial
saddle points. This program can be definitely completed in one dimension.
First of all set 
\begin{eqnarray}
f(r)= \frac{\dot{y}}{y}
\end{eqnarray}
Obviously, 
\begin{eqnarray}
\frac{\ddot{y}}{y}=\dot{f}+f^2
\end{eqnarray}
Let us now rewrite the functional in Eq.(\ref{cipolla}) in terms of $f(r)$.
We obtain 
\begin{eqnarray}
I[f(r)]=\frac{1}{2}\int\;dr\;\left(\dot{f}+f^2-2E\right)^2
\end{eqnarray}
Using the Euler-Lagrange equation, one may show that the nontrivial saddle
point is given by the solution of 
\begin{eqnarray}
\ddot{f}-2f^3+4\;E\;f=0
\end{eqnarray}
This equation describes the motion in an inverted double well potential with
maxima at $f_{\pm}=\pm\sqrt{2E}$. The corresponding instanton is a
trajectory going from $f_{+}$ to $f_{-}$. Integrating this equation one
obtains 
\begin{eqnarray}
f(r)=-\sqrt{2E}\tanh(\sqrt{2E}r)
\end{eqnarray}
Correspondingly we find 
\begin{eqnarray}
y_0(r) \propto \mathrm{sech}(\sqrt{2E}x)
\end{eqnarray}
Finally let us evaluate the action corresponding to the potential relative
to $y_0(r)$. We have to calculate $S=1/(2\gamma)I[y_0(r)]$. Inserting the
expression for $y_0$ and evaluating the integral one obtains 
\begin{eqnarray}
S=\frac{2}{3}\frac{1}{\gamma}(2E)^{3/2}
\end{eqnarray}
This is corresponds to the action obtained by Halperin in its exact solution
of the 1d problem, Ref.[\onlinecite{Halperin}].

\section{Gaussian fluctuations.}

In this appendix we outline the calculation of the full expression of the
subgap tails of the DOS, including gaussian fluctuations.

For every function $f(r)$, we select the center of the $r^{\prime}$ of the
optimal fluctuation in such a way as to minimize the functional 
\begin{eqnarray}  \label{functional}
D(f\mid r^{\prime})=\int dr \left[f(r)-f_0(r-r^{\prime}) \right].
\end{eqnarray}
Of all the solutions, we select the $r^{\prime}$ closest to $r$, the point
at which the DOS is evaluated. Subsequently, the function $f(r)$ is expanded
in an complete set 
\begin{eqnarray}  \label{expansion}
f(r)=\sum_{n}^{} \xi_n \phi_n(r),
\end{eqnarray}
where the first $d+1$ functions of the expansion are selected according to 
\begin{eqnarray}
\phi_0(r)&=&a\;f_0(r-r^{\prime}), \\
\mathbf{\phi}_1(r)&=&b \nabla f_0(r-r^{\prime}).
\end{eqnarray}
Here the constant $a$ and $b$ are given by 
\begin{eqnarray}  \label{a}
a&=& \left[ \int dr (f_0(r))^2 \right]^{1/2}, \\
b&=& \left[\frac{1}{d} \int dr (\nabla f_0(r))^2 \right]^{-1/2}.  \label{b}
\end{eqnarray}

Having in mind this expansion, we write 
\begin{eqnarray}
&&\langle y(r) \rangle= \langle \int_{\sigma} dr^{\prime\prime} y(f(r))
\delta(r^{\prime\prime}-r^{\prime})\rangle  \nonumber \\
&&= \langle \int_{\sigma} dr^{\prime\prime} y(f(r)) \delta(\nabla D(f\mid
r^{\prime\prime})) \mid \mathrm{det}\nabla \nabla D \mid \rangle,
\end{eqnarray}
where $\sigma$ is a region of space containing at most a single optimal
fluctuation, $y(f(r))$ the appropriate solution of 
Eq.(\ref{saddlepoint1})-(\ref{saddlepoint2}) 
relative to $f(r)$, and in the last line we operated a change of variables
from $r^{\prime\prime}$ to $D$.

Using Eq.(\ref{functional})-(\ref{expansion}) we can estimate 
\begin{eqnarray}
\nabla D&=& \frac{2 \mathbf{\xi}_1}{b}, \\
\mid {\det \nabla \nabla D} \mid &\simeq& \frac{2^d}{b^{2d}}.
\end{eqnarray}
At the same time, while to lowest order $y(f(r)) \propto y_0(r-r^{\prime})$,
the coefficient of proportionality is set by fluctuations of $f$ around $f_0$%
. Using Eq.(\ref{saddlepoint1}), Eq.(\ref{saddlepoint2}), 
together with Eq.(\ref{expansion}), one obtains 
\begin{eqnarray}
y(f(r)) &\approx& A\;y_0(r), \\
A&=& \sqrt{\frac{\xi_0-a^{-1}}{\int dr \phi_0(r)[y_0(r)]^2}}.
\end{eqnarray}

Now using the results above, it is easy to evaluate 
\begin{eqnarray}  \label{finalalmost}
&&\langle y \rangle \simeq \frac{1}{b^d} \langle \int dr^{\prime\prime}
y(f(r)) \delta(\mathbf{\xi}_1) \rangle  \nonumber \\
&&= \frac{1}{(2\pi \eta)^{3/2}\;b^d}\int dr^{\prime\prime} \frac{d\xi_0}{%
(2\pi \eta)^{1/2}} \;A\;y_0 e^{-\frac{(\xi_0)^2}{2\eta}}  \nonumber \\
&&= \frac{I_0}{(2\pi \eta)^{2}}\int_{a^{-1}}^{+\infty} d\xi_0 \sqrt{\xi_0
-a^{-1}}\;e^{-\frac{(\xi_0)^2}{2\eta}},
\end{eqnarray}
where 
\begin{eqnarray}  \label{I0}
I_0=\frac{\int dr y_0(r)}{\sqrt{\int dr \phi_0(r) [y_0(r)]^2}}\;\frac{1}{b^d}%
.
\end{eqnarray}
The last object we have to evaluate is the integral describing the
contribution to the DOS due to gaussian fluctuations of the height of the
optimal fluctuation, i.e. 
\begin{eqnarray}  \label{I}
I&=&\int_{a^{-1}}^{+\infty} d\xi_0 \sqrt{\xi_0 -a^{-1}}\;e^{-\frac{(\xi_0)^2%
}{2\eta}}  \nonumber \\
&=&\eta^{3/4}\Gamma(3/2) \exp[-S/2]\;D_{-3/4}(\sqrt{2\;S})  \nonumber \\
&\simeq& \eta^{3/4}\Gamma(3/2)\left(\frac{1}{2S}\right)^{3/4} e^{-S}
\end{eqnarray}
where $S$ is the action at the saddle point calculated before, $D_{-3/4}$ is
a parabolic cylinder function, and the last line is the lowest order
asymptotic expansion for $S \gg 1$.

Using now Eq.(\ref{finalalmost})-(\ref{I0})-(\ref{I}), and Eq.(\ref{DOS1})
one obtains the final result Eq.(\ref{finalDOS}).


\section{Estimate of the Qubit relaxation rate}

~\label{appc}

In this Appendix, we estimate roughly the Qubit relaxation rate due to the
interaction between a single localized quasiparticle and the phase across
the Josephson junction. In the following, we will have in mind large
Josephson junctions of area $A\approx 10\times 10\;\mathrm{\mu m}$ [see
Fig.~(\ref{fig:fig1})] with electrodes of thickness $l_{z}\approx 0.1\;{\mu m}$,
operating at a linear frequency of $f=10\;\mathrm{GHz}$, corresponding to a
circular frequency of $\omega _{0}\approx 6\;10^{10}\;\mathrm{s^{-1}}$.

Let us consider the phase $\varphi$ across the junction evolving in time
according to $\varphi(t)=\cos(\omega_0 t)$. Correspondingly, the voltage $V$
across the junction will evolve as $V(t)=\hbar/2e\;\dot{\varphi}$. For
typical junctions, having oxide layers of thickness $~\approx 10\;\mathring{A%
}$, the potential drop is located mostly in a thin layer of width $d\approx
1\;\mathring{A}$ within the superconducting electrodes. The rate at which a
quasiparticle localized on a region of linear size $\lambda \approx 1\;{\mu m%
}$ is excited due its interaction with this time dependent potential, can be
estimated as 
\begin{eqnarray}
\Gamma_{qp}\simeq \frac{2\pi }{\hbar} \sum_{f} \mid \int dr \psi_{i}(r)\;
eV(r)\; \psi^{*}_{f}(r) \mid^2  \nonumber \\
\delta(\epsilon_{f}-\epsilon_{i}-\hbar \omega_0),
\end{eqnarray}
where $\psi_{i,f}$ and $\epsilon_{i,f}$ are the wave functions and energies
of the initial and final state respectively. Using the fact that the
interaction is mostly concentrated close to the surface $S$ of the
electrode, we can write 
\begin{eqnarray}
~ \Gamma_{qp} \simeq \frac{\pi}{2\hbar} (\hbar\omega_0)^2 \left(\frac{d}{l_z}%
\right)^2\;\gamma(\omega_0),  \label{gammaqp}
\end{eqnarray}
where 
\begin{eqnarray}
\gamma \simeq l_z^2 \sum_{f} \mid \int_{S} dr \psi_{i}(r) \psi^{*}_{f}(r)
\mid^2 \delta(\epsilon_{f}-\epsilon_{i} - \hbar \omega_0).
\end{eqnarray}
Since the support of the wave function of the initial state is concentrated
on an area $A \simeq \lambda\times\lambda \approx 1 \times 1\;{\mu m^2}$, $%
\gamma$ can be roughly estimated having in mind a three dimensional
diffusive quantum dot of dimensions $\lambda \times \lambda \times l_z$ as 
\begin{eqnarray}  \label{gamma}
\gamma \approx \frac{ l_z }{ 4\pi^2 \nu_0\;\lambda^2}\;\int_S\;dr dr^{\prime}%
\left[G(r,r^{\prime},\epsilon)\right]_{-} \left[G(r^{\prime},r,\epsilon+%
\hbar\omega_0)\right]_{-},
\end{eqnarray}
where $\nu_0$ is the DOS per unit volume, and 
\begin{eqnarray}
[G]_{-}=-i(G^a-G^r),
\end{eqnarray}
$G^{r,a}$ being the QD's retarded/advanced Green's functions.

Performing a disorder average~\cite{Aleiner2} on Eq.(~\ref{gamma}), one
arrives at 
\begin{eqnarray}
\Gamma_{qp} \simeq \hbar \omega_0^2 \left(\frac{d}{l_z}\right)^2\left[
(\nu_0 \lambda^2 l_z)\; \left(\frac{\lambda_F}{\lambda}\right)^2 + \frac{1}{6%
} \frac{\;l_z^2}{\hbar D} \right],
\end{eqnarray}
where $D$ is the diffusion constant. Finally, inserting this expression into
Eq.(\ref{gammaqp}), and using the typical parameters reported above [$%
\lambda_F \approx 1\;\mathring{A},\; D \approx 50 \mathrm{cm^2/s},\; \nu_0
\approx 1.5\;10^{22}\;\mathrm{cm^{-3}\;eV^{-1}}$], one arrives at $%
\Gamma_{qp} \approx 10^3 \;\mathrm{Hz}$.


\begin{thebibliography}{99}
\bibitem{Mohanty} P. Mohanty, E. M. Q. Jariwala and R. A. Webb, Phys. Rev.
Lett. \textbf{78}, 3366 (1997).\textbf{\textrm{\ }}

\bibitem{Pierre} F. Pierre and N. O. Birge, Phys. Rev. Lett.\textbf{\ 89},
206804 (2002).

\bibitem{Pierre2} F. Pierre, A. B. Gougam, A. Anothore, H. Pothier, D.
Esteve, and N. O. Birge, Phys. Rev. B \textbf{68} , 085413 (2003).

\bibitem{Glazman} A. Kaminsky and L. I. Glazman, Phys. Rev. Lett. 86 , 2400
(2001).

\bibitem{Pothier} H. Pothier, S. Gueron, N. O. Birge, D. Esteve, and M. H.
Devoret, Phys. Rev. Lett. \textbf{79} , 3490 (1997).

\bibitem{Pierre3} F. Pierre, H. Pothier, D. Esteve, M. H. Devoret, A.
Gougam, and N. O. Birge, in \textit{Kondo Effect and Dephasing in
Low-Dimensional Metallic Systems,}

edited by V. Chandrasekar, C. Van Haesendonck and A. Zawadowsky, (Kluwer,
Dordrecht, 2001) ; cond-mat/0012038.

\bibitem{Abrikosov} A. A. Abrikosov, and L. P. Gorkov, JETP \textbf{12 },
1243 (1961).

\bibitem{Balatsky} A. V. Balatsky, and S. A. Trugman, Phys. Rev. Lett 
\textbf{79}, 3767 (1997).\textbf{\textrm{\ }}

\bibitem{Aleiner} I. S. Beloborodov, B. N. Narozhny, and I. L. Aleiner,
Phys. Rev. Lett. \textbf{85}, 816 (2000).

\bibitem{Shytov} A. V. Shytov, I. Vekhter, I. A. Gruzberg, and A. V.
Balatsky, Phys. Rev. Lett. \textbf{90} , 147002 (2003); I. Vekhter, A. V.
Shytov, I. A. Gruzberg, A. V. Balatsky, Physica B\textbf{\ 329-333} , 1446
(2003).

\bibitem{Brouwer} M. G. Vavilov, P. W. Brouwer, V. Ambegaokar, and C. W. J.
Beenakker, Phys. Rev. Lett. \textbf{86} , 874 (2001).

\bibitem{Lamacraft} A. Lamacraft, and B. D. Simons, Phys. Rev. Lett. \textbf{%
85} , 4783 (2000); A. Lamacraft, and B.D. Simons, Phys. Rev. B \textbf{64},
014514 (2001).

\bibitem{Martinis1} K. M. Lang, S. Nam, J. Aumentado, C. Urbina, and J. M.
Martinis, IEEE Transactions on Applied Superconductivity \textbf{13}, 989
(2003).

\bibitem{Anderson} P. W. Anderson, J. Phys. Chem. Solids \textbf{11}, 26
(1959).

\bibitem{Liu} Lu Yu, Phys. Sin. \textbf{21} , 75 (1965).

\bibitem{Rusinov} A. I. Rusinov, JETP Lett. 9 , 85 (1969).

\bibitem{Balatsky3} M. I. Salkola, A. V. Balatsky, and J. R. Schrieffer,
Phys. Rev. B 55 , 12648 (1997).

\bibitem{Shiba} H. Shiba, Progr. Theor. Phys. 40 , 435 (1968).

\bibitem{Rusinov2} A. I. Rusinov, Sov. Phys. JETP, 29 , 1101 (1969).

\bibitem{Lifshits} I. M. Lifshits, Sov. Phys. Usp. 7 , 549 (1965).

\bibitem{Halperin} B. Halperin, Phys. Rev. 139 , A104 (1965).

\bibitem{Lax} B. Halperin and M. Lax, Phys. Rev. 148 , 722 (1966).

\bibitem{Zittarz} J. Zittarz and J. S. Langer, Phys. Rev. 148 , 741 (1966).

\bibitem{Larkin} L. D. Larkin and Yu. N. Ovchinikov, JETP 34 , 1144 (1972).

\bibitem{Tracy} C. A. Tracy, and H. Widom, Comm. Math. Phys. 159 , 151
(1994); 177 , 727 (1996).

\bibitem{Balatsky1} A rather extreme limit of this phenomenon was considered
in Ref.[\onlinecite{Balatsky}], where the authors studied the effect of
fluctuations of the concentration of magnetic impurities exceeding the AG
critical concentration $n_{c}$, and therefore generating gapless regions
optimal fluctuations surrounded by a gap-full matrix.

\bibitem{Usadel} K. Usadel, Phys. Rev. Lett. 25 , 507 (1970).

\bibitem{Kopnin} N. Kopnin, Theory of Nonequilibrium Superconductivity,
(Clarendon Press, Oxford, 2001).

\bibitem{Meyer} J. S. Meyer, and B. D. Simons, Phys. Rev. B 64 , 134516
(2001).

\bibitem{Ostrovsky} P. M. Ostrovsky, M. A. Skvortsov, M. V. Feigel'man,
Phys. Rev. Lett 87 , 027002 (2001).

\bibitem{Dima} D.E. Khmelnitskii, in New Directions in Mesoscopic Physics
(Towards Nanoscience), (NATO ASI, Edt. R. Fazio, V. F. Gantmakher, and Y.
Imry, Kluwer, Dordrecht, 2003).

\bibitem{foot} The Euler-Lagrange equation Eq.(\ref{NonTrivialSaddle}) is of
fourth order in the gradients of $x$, while the equation we solve, Eq.(\ref%
{EL1}), is of second order. By reducing the order of the equations, some
solutions could be lost. In order to prove that the analytical solution that
we found is indeed the optimal fluctuation, we minimized numerically the
action Eq.(\ref{actionmin}) in three dimensions, using a variational ansatz
with two free parameters. The numerical result for the constant $\tilde{a}%
_{d}$ [see Eq.(\ref{3d}), and Eq.(\ref{finalDOS})] is $~64$, within 2
percent of the result obtained analytically.

\bibitem{Nakamura} Y. Nakamura, Yu. A. Pashkin, and J. S. Tsai, Nature 398 ,
786 (1999).

\bibitem{Mooji} C. H. van der Wal, A. C. J. ter Haar, F. K. Wilhelm, R. N.
Schouten, C. J. P. M. Harmans, T. P. Orlando, S. Lloyd, and J. E. Mooij
Science 290 , 773 (2000).

\bibitem{Devoret} D. Voin, A. Aassime, A. Cottet, P. Joyez, H. Pothier, C.
Urbina, D. Esteve, and M. H. Devoret, Science 296, 886 (2002); see also A.
Cottet, Ph. D. Thesis (Paris, 2002, unpublished).

\bibitem{Martinis2} J. H. Martinis, S. Nam, J. Aumentado, and C. Urbina,
Phys. Rev. Lett. 89, 117901 (2002).

\bibitem{Burin1995} A. Burin, J. Low. Temp. Phys. \bf 100 \rm, 309 (1995).

\bibitem{Gershenson} M. Gershenson, private communication.

\bibitem{noteA} Notice that, in contrast with the case of Lifshits tails, in
the present situation each optimal fluctuation contains typically a number $%
N\gg 1$ of quasiparticle states.

\bibitem{tau} Indeed, we expect $\tau _{tr}$ to be of the same order of
magnitude as the phonon limited quasiparticle lifetime $\tau _{ph}$ at the
Fermi level in Al. In the diffusive limit~\cite{Lawrence}, the latter can be
estimated to be approximately to be $1/\tau _{ph}\simeq \lambda \pi
^{4}/2\;(k_{B}T/\omega _{D})^{3}(k_{F}l)(k_{D}/k_{F})^{3}\;k_{B}T/\hbar $,
where $k_{D}$ and $\omega _{D}$ are the Debye momentum and frequency, $k_{F}$
is the Fermi momentum, $l$ is the mean free path, and $\lambda $ the
electron phonon coupling constant. At $T=10\;mK$, using $\lambda \approx 1$,$%
\omega _{D}\approx 1000\;K$, $k_{D}/k_{F}\approx 1$, and $k_{F}l\approx
10^{2}$ one obtains $\tau _{ph}\approx 10^{2}\;s$.

\bibitem{Lawrence} A. Sergeev and V. Mitin, Phys. Rev. B 61 , 6041 (2000).

\bibitem{Aleiner2} I.L. Aleiner, P.W. Brouwer, L.I. Glazman, Phys. Rep. 358
, 309 (2002).
\end{thebibliography}
\end{document}